\documentclass[aps,prl,reprint,superscriptaddress, showpacs]{revtex4-1}

\usepackage{graphicx}
\usepackage{natbib}
\usepackage[english]{babel}
\usepackage{amsmath}
\usepackage{units}
\usepackage{color}
\usepackage{pifont}

\begin{document}

\title{Long-range Synchronization of Nanomechanical Oscillators with Light}

\author{Shreyas Y. Shah}
\affiliation{School of Electrical and Computer Engineering, Cornell University, Ithaca, New York 14853, USA}
\affiliation{School of Electrical and Computer Engineering, Purdue University, West Lafayette, Indiana 47907, USA}
\author{Mian Zhang}
\affiliation{School of Electrical and Computer Engineering, Cornell University, Ithaca, New York 14853, USA}
\affiliation{School of Engineering and Applied Sciences, Harvard University, Cambridge, Massachusetts 02138, USA}
\author{Richard Rand}
\affiliation{Department of Mathematics, Cornell University, Ithaca, New York 14853, USA}
\affiliation{Sibley School of Mechanical and Aerospace Engineering, Cornell University, Ithaca, New York 14853, USA}
\author{Michal Lipson}
\affiliation{School of Electrical and Computer Engineering, Cornell University, Ithaca, New York 14853, USA}
\affiliation{Kavli Institute at Cornell for Nanoscale Science, Ithaca, New York 14853, USA}
\affiliation{Department of Electrical Engineering, Columbia University, New York, New York 100027, USA}

\begin{abstract}
We experimentally demonstrate mutual synchronization of two free-running nanomechanical oscillators separated by an effective distance of 30 meters and coupled through light. Due to the finite speed of light, the large separation introduces a significant coupling delay of 139 nanoseconds, approximately four and a half times the mechanical oscillation time period. We reveal multiple stable states of synchronized oscillations, enabled by delayed coupling, with distinct synchronization frequency in the coupled system. These states are accessed by tuning independently the directional coupling strengths. Our results demonstrate rich dynamics and could enable applications in reconfigurable radio-frequency networks and novel computing concepts.
\end{abstract}


\maketitle

Synchronization is of fundamental importance in natural systems such as neuronal networks \cite{deco2009, dhamala2004}, chemical reactions \cite{erneux2008} and biochemical systems \cite{takamatsu2000, takamatsu2006, rossoni2005, kauffman1975}. Synchronization in micro- and nanomechanical oscillators promises applications in communication  \cite{bregni2002}, signal-processing \cite{stephan1986} and novel complex networks \cite{hoppensteadt2001, fischer2005}. Synchronizing nanoscale mechanical oscillators have recently been achieved on various platforms including nano-electromechanical systems and optomechanical systems \cite{shim2007, zhang2012, hoppensteadt2001, milburn2012, milburn2014, cross2004, zhang2015, bagheri2013, matheny2014, seshia2013, heinrich2011}. So far these demonstrations of synchronization are only achieved on nanomechanical oscillators separated by a short distance due to restrictions imposed by their coupling mechanism. Mechanical coupling \cite{shim2007}, optical evanescent coupling \cite{zhang2012,zhang2015} and coupling through a common optical cavity \cite{bagheri2013} mode are restricted to small separations of micrometers due to high loss and low efficiency coupling at long distances. Therefore a major challenge in realizing a synchronized network of nanomechanical oscillators is to be able to mutually couple them at a long distance, many orders of magnitude greater than the size of each oscillator, in a controllable fashion. 

\begin{figure}[h!]
\includegraphics[scale=0.2]{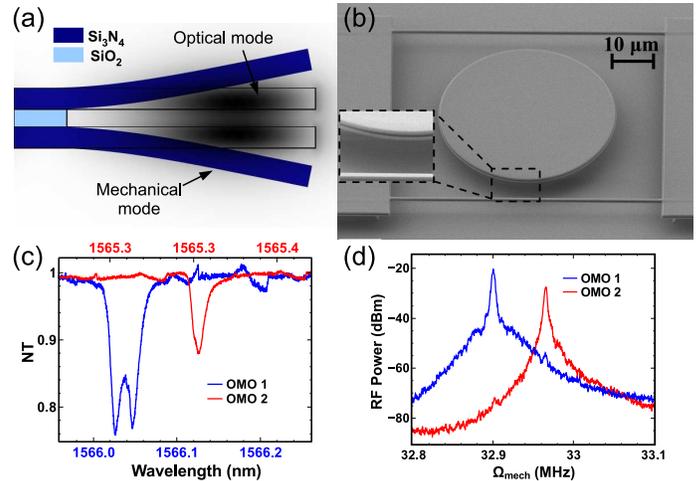}
\caption{\textbf{(a)} Schematic cross-sectional picture of the periphery of a typical optomechanical (OM) resonator, indicating the co-localization of optical and mechanical modes. \textbf{(b)} SEM image of a typical double-disk OM resonator, surrounded by a structure to support tapered optical fibers used to optically excite mechanical oscillations. \textbf{(Inset)} Higher-magnification image of the double microdisk structure. \textbf{(c)} Normalized optical transmission spectra of the two OM resonators used in this demonstration. The split in the optical resonance of OMO 1 is due to scattering from imperfections, that couples the clockwise and counter-clockwise modes. \textbf{(d)} RF power spectrum of the transmitted optical power, modulated by each device as it is driven into self-sustained oscillations by a cw laser. The oscillation frequencies are separated by 70 kHz.}
\label{fig1}
\end{figure}

In this paper, we demonstrate mutual synchronization of two nanomechanical oscillators separated by effectively 28.5 m using light guided through optical fibers that introduce minimal loss over a long-distance. The separation between the two oscillators is six orders of magnitude larger than the individual oscillating cantilevers, which themselves are only microns long. Such long separation introduces significant time-delay in the coupling that is approximately 4.5 times of the oscillation time period of each individual oscillator. We show that with the presence of this significant separation and time-delay, the two nanomechanical oscillators can still synchronize and display a multitude of stable states of synchronized oscillations that are unique in time-delayed systems \cite{choi2000, kim1997, schuster1989, fischer2005, takamatsu2000, takamatsu2006}. In contrast to synchronization in systems that have similar bidirectional coupling strengths \cite{zhang2012, zhang2015, bagheri2013, matheny2014, seshia2013}, we demonstrate independent tuning of the coupling strength in each direction and show the emergence of new synchronization frequencies.

Each oscillator used in the experiment has a double microdisk structure (Fig. \ref{fig1}) that supports coupled optical and mechanical resonances, and is driven into self-sustained, free running oscillations with an external laser. The double microdisk structure is composed of two vertically stacked suspended microdisks with a spacer between them (Fig. \ref{fig1}(a)). The top and bottom disks are made of low-pressure chemical vapor deposition (LPCVD) grown silicon nitride (Si$_3$N$_4$), and are 250 nm and 220 nm thick respectively. The spacer is made of 170 nm thick plasma-enhanced CVD grown silicon dioxide (SiO$_2$). This stack rests on a 4 $\mu$m thick substrate of thermally grown SiO$_2$. These thin films are patterned into disks with a 20 $\mu$m radius using electron-beam lithography and inductively coupled reactive ion etching. The SiO$_2$ layers are partially etched away with buffered hydrofluoric acid to release the periphery of the disks (Fig. \ref{fig1}(b)). The resulting structure supports a high-quality-factor whispering gallery optical mode that is coupled to the antisymmetric mechanical vibration mode of the two freestanding edges (Fig. \ref{fig1}(a)). The coupling strength between the optical and mechanical degrees of freedom is characterized by the optomechanical coupling constant $G_{\textrm{om}}$= -$2\pi\times 45$ GHz/nm, as deduced from finite element simulations \cite{zhang2012}. 

The two oscillators are coupled by using the radio-frequency (RF) oscillations of one oscillator to modulate the optical drive of the other oscillator. Each optomechanical oscillator (OMO) can be modeled as a mechanical oscillator (Eq. \ref{eqn1}) with natural frequency $\Omega_i$ and damping rate $\Gamma_i$. The oscillators are driven into spontaneous self-sustained oscillations by two independent continuous wave (cw) lasers. When the optical coupling is off, each oscillator vibrates close to its natural frequency and produces an intensity modulation on its respective cw pump laser. Due to the nature of the self-feedback process of optomechanical oscillators, the phase of the oscillations is completely free. When the optical coupling is on, the drive laser intensity of one OMO is modulated by the mechanical displacement signal of the other OMO $x_j(t-T)$ (and vice-versa) after a propagation delay of $T$ and a coupling-constant $\gamma_{ij}$. This coupling ultimately leads to the synchronization between the two OMOs. The governing equation of each OMO $i$ is expressed as

\begin{equation}\label{eqn1}
\begin{split}
\ddot{x}_{i}(t) + &\Gamma \dot{x}_{i}(t) + \Omega_{i}^2 x_{i}(t) = \\
&F_{opt \text{ }i}(x_{i}(t))[1+\gamma_{ij}f(x_{j}(t-T))]
\end{split}
\end{equation}

where $f(x_{j}(t-T)$ is a normalized forcing function that transduces the delayed modulation from the other OMO $j$ \cite{shah2015}.

\begin{figure*}[ht!]
\includegraphics[width=\textwidth]{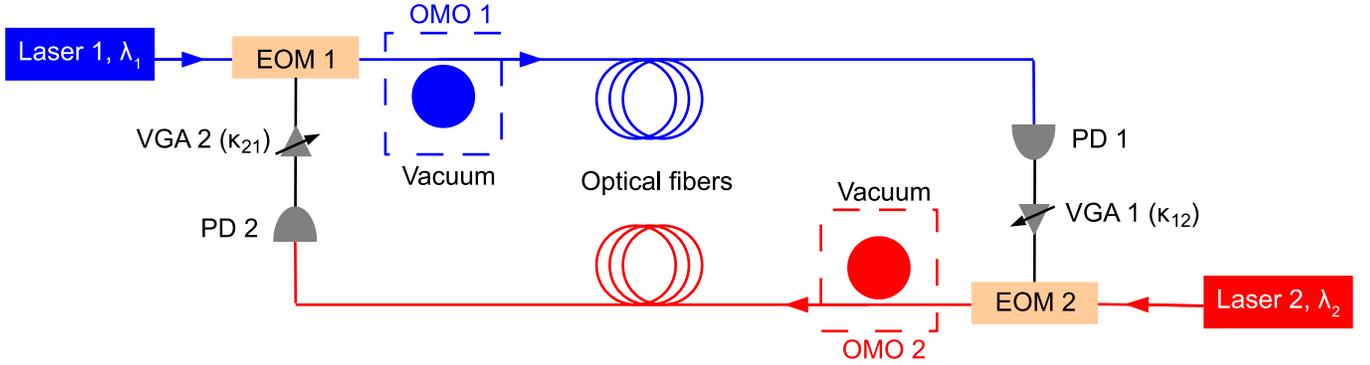}
\caption{Schematic of the experimental setup to synchronize two optomechanical oscillators (OMOs). Each device is driven by an independent laser tuned to the blue-side of its optical resonance. The transmitted optical signals, modulated by each OMO, travel over a 9 m long delay line of optical fibers. The RF signal generated at the photodetector (PD) at the end of optical delay line 1 modulates the power of the laser driving OMO 2 via an electro-optic modulator (EOM), and vice-versa. The strengths of these modulation signals are controlled by variable-gain RF amplifiers (VGA). Half of the RF oscillation signal is tapped off at each of the photodetectors for analyzing with an RF spectrum analyzer (See Supplemental Material \cite{supplementary2017} for a detailed schematic).}
\label{fig2}
\end{figure*}

We achieve long-range mutual coupling of two OMsupplemOs, with individual mechanical oscillation frequencies of 32.90 MHz and 32.97 MHz, over an effective distance of 28.5 m (delay $T=139$ ns) via low loss optical fibers (Fig. \ref{fig2}). The RF oscillations of OMO 1 (Fig. \ref{fig2}) are carried by light over the optical fiber, and converted to an electrical signal using a high-speed photodetector. This electrical signal modulates the power of the laser driving OMO 2 using an electro-optic modulator, EOM 2, thereby coupling OMO 1 to OMO 2. Similarly, OMO 2 couples to OMO 1 via EOM 1. 

The EOMs enable independent control of the directional coupling strengths. The optical coupling strengths are controlled with variable-gain RF amplifiers (VGA 1,2), that can provide an arbitrary gain between -26 dB and +35 dB. The coupling strengths $\kappa_{ij}$,($i,j = 1,2$) are defined as  $\kappa_{ij} = H_{\textrm{in},i}/H_{\textrm{osc},j}$, where $H_{\textrm{osc},j}$ is the oscillation power of OMO $j$, and $H_{\textrm{in},i}$ is the power of the signal from OMO $i$ imparted on the laser via EOM $j$ driving OMO $j$. It can be shown that $\kappa_{ij}$ is proportional to $\gamma_{ij}$ from Eq. 1 (see Supplemental Material \cite{supplementary2017}). As depicted in Fig. 2, $\kappa_{12}$ is controlled by VGA 1 and $\kappa_{21}$ is controlled by VGA 2. We also introduce a third VGA to control the overall coupling strength while keeping the ratio $\kappa_{12}/\kappa_{21}$ is fixed (not shown in schematic in Fig. \ref{fig2}).

\begin{figure*}[ht!]
\includegraphics[scale=0.35]{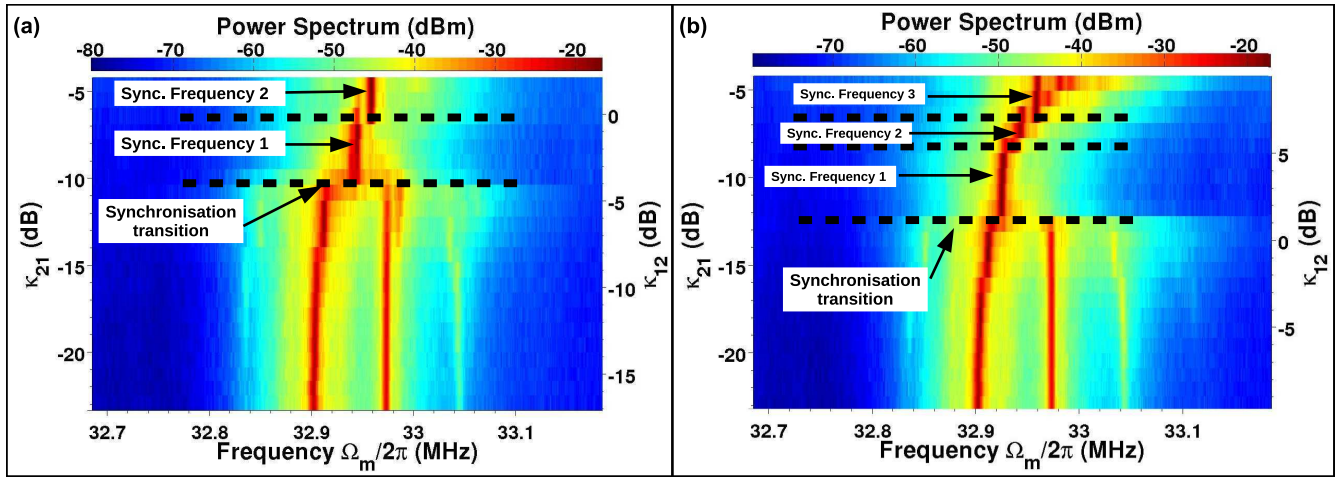}
\caption{Combined power spectrum of the transmitted optical power of the two OMOs, as a function of increasing coupling strengths $\kappa_{21}$ ($\kappa_{12}$), showing synchronization transitions, while $\kappa_{12}/\kappa_{21}$ is kept constant at \textbf{(a)} 6.32 dB and \textbf{(b)} 13.63 dB. As $\kappa_{21}$ and $\kappa_{12}$ are increased beyond the synchronization threshold, we see \textbf{(a)} 2 synchronized states and \textbf{(b)} 3 synchronized states, respectively.}
\label{fig3}
\end{figure*}

We demonstrate the onset of long-range synchronization of the two oscillators, as they transition from oscillating independently as we increase the coupling strength. When the oscillators are weakly coupled (small values of $\kappa_{ij}$), they oscillate at their individual frequencies (Fig. \ref{fig3}). As the coupling strength is increased, we observe frequency-pulling  \cite{cross2004, razavi2004} i.e. the frequencies of the two oscillators are pulled towards each other while they still oscillate independently, prior to the onset of synchronized oscillations. As the coupling strength is increased beyond a threshold value, $(\kappa_{12},\kappa_{21}) = (-10.5 ~\textrm{dB}, -4.1~\textrm{dB})~\textrm{and}~(-12.1~\textrm{dB}, 1.5~\textrm{dB})$ as shown in Fig. \ref{fig3}(a) and Fig. \ref{fig3}(b) respectively, the two oscillators spontaneously begin to oscillate in synchrony, as indicated by the emergence of a single RF tone in the transmitted optical power spectrum.

We show that the frequency of long-range synchronization can be controllably changed by changing the overall coupling strength. The optical propagation time between the two oscillators imposes a significant delay (relative to the oscillation time period) in the coupling dynamics. Several theoretical studies  \cite{matsumoto1987,schuster1989, strogatz1999, choi2000, kim1997, rand2002, yanchuk2005, julicher2014, yanchuk2015} and a few experimental works \cite{fischer2005,takamatsu2000,takamatsu2006,erneux2009} on delay-coupled oscillators show that multiple frequencies of stable synchronized oscillations can be possible in such systems. This is because multiple stable oscillation states are possible with a single value of delay. Intuitively, the multiplicity can be seen as arising from the interactions between the (multiple) resonances of the coupling feedback loop and the oscillators \cite{schuster1989,yanchuk2005}. Indeed, these OMOs exhibit multiple stable states in which they oscillate synchronously (Fig. 3), as opposed to just a single stable synchronized state seen in systems without delay \cite{zhang2012, zhang2015, bagheri2013, matheny2014, seshia2013}. As shown in Fig. 3(a), for a fixed ratio of the two-way coupling ($\kappa_{12}/\kappa_{21}$ = 6.3 dB), the two oscillators change from a synchronized $(\kappa_{21}=-10.5$ dB) state at 32.93 MHz to a second synchronized state at 32.94 MHz when the overall coupling strength is increased by $\sim$ 4dB ($\kappa_{21}=-6.5$ dB). 

We show that the discrete frequencies of synchronization can be accessed by varying the relative directional coupling strength $\kappa_{12}$ and $\kappa_{21}$ individually \cite{gielen2009}. Fig. \ref{fig3}(b) shows an extra frequency accessible for the same range of values for $\kappa_{21}$ when the coupling strength ratio is tuned to $\kappa_{12}/\kappa_{21}=13.63$ dB. We choose a few coupling strength ratios and show a ladder of accessible synchronized frequencies in Fig. 4 with various two-way coupling ratios. The discrete points in Fig. 4 represent the frequency of the synchronized states for different values of the ratio $\kappa_{12}/\kappa_{21}$ and they are plotted versus the value of $\kappa_{21}$ at which they occur. The value of the synchronized frequency $\Omega_{\textrm{sn}}$ is normalized to the frequency difference of the two oscillators $\Omega_1$ and $\Omega_2$. For a very small value of the ratio $\kappa_{12}/\kappa_{21}$ i.e. when OMO 2 dominates the interaction between the two oscillators, the two OMOs synchronize at $\Omega_{\textrm{sn}} = \Omega_2$. This is similar to master-slave locking \cite{shah2015}, where the coupling is unidirectional from one oscillator to the other. In a more evenly coupled scenario, with $\kappa_{12}/\kappa_{21}=6.3$, the first onset of synchronization is at $\Omega_{\textrm{sn}} = 0.65$ , close to their average frequencies. The synchronised oscillations then transition to $\Omega_{\textrm{sn}} = 0.85$ at a stronger overall coupling strength. With intermediate coupling ratios of $\kappa_{12}/\kappa_{21}=$11.9 and 13.6 dB (purple and red), a new synchronization frequency occurs at $\Omega_{\textrm{sn}} = 0.4$ which is inaccessible in the scenario with $\kappa_{12}/\kappa_{21}=6.3$.  Further changes in the coupling ratio could lead to synchronized states with frequencies that span the entire range between the two natural frequencies in discrete steps.

\begin{figure}[h!]
\centering
\includegraphics[scale=0.42]{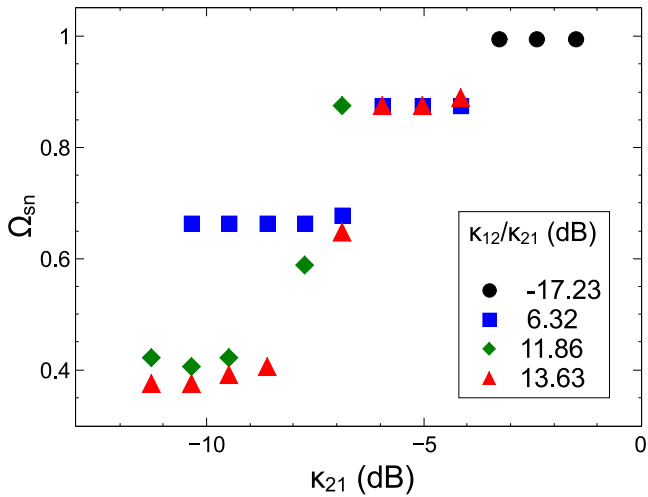}
\caption{Synchronization frequency (normalised), $\Omega_{\textrm{sn}} = \frac{\Omega_{\textrm{sync.}} - \Omega_{\textrm{1}}}{\Omega_{\textrm{2}}-\Omega_{\textrm{1}}}$, as obtained by varying $\kappa_{21}$, with different values of the ratio $\kappa_{12}/\kappa_{21}$. It can be seen that, for the values of $\kappa_{21}$ and $\kappa_{12}$ used in this demonstration, $\Omega_{\textrm{sn}}$ takes a value in one of four discrete clusters. As we show in Fig. 3(a) (corresponding to blue squares) and Fig. 3(b) (corresponding to red triangles), oscillation states which are not available for one particular ratio of directional coupling constants can be accessed by changing varying the value of $\kappa_{12}/\kappa_{21}$.}
\label{fig4}
\end{figure}

Our demonstration of long-range, controllable and multi-stable synchronization between delay-coupled OMOs paves a way towards practical implementation of scalable networks of distributed nanomechanical oscillators. Independent control of the directional coupling strength enables us to choose from multiple possible synchronization states with different oscillation frequencies. Such networks could enable applications of OMOs in distributed, reconfigurable communication networks and novel memory concepts \cite{bregni2002,hoppensteadt2001}. Delayed-coupling also manifests itself in biological systems \cite{takamatsu2000,takamatsu2006,kauffman1975,julicher2012}, particularly neuronal networks. Delay-coupled OMOs can potentially serve as a platform to implement various schemes of neuromorphic information processing and computation \cite{hoppensteadt2001,lee2013,cosp2004}.

We acknowledge Lauren Lazarus for useful discussions on theoretical aspects of coupled nonlinear oscillators. The authors also gratefully acknowledge support from DARPA for award W911NF-11-1-0202 supervised by Dr. Jamil Abo-Shaeer. The authors also acknowledge Applied Optronics. This work was performed in part at the Cornell NanoScale Facility, a member of the National Nanotechnology Coordinated Infrastructure, which is supported by the National Science Foundation (Grant ECCS-15420819). This work made use of the Cornell Center for Materials Research Facilities supported by the National Science Foundation under Award Number DMR-1120296. The authors also acknowledge Prof. Paul McEuen for use of lab facilities.

\bibliography{LDFS}

\begin{thebibliography}{38}%
\makeatletter
\providecommand \@ifxundefined [1]{%
 \@ifx{#1\undefined}
}%
\providecommand \@ifnum [1]{%
 \ifnum #1\expandafter \@firstoftwo
 \else \expandafter \@secondoftwo
 \fi
}%
\providecommand \@ifx [1]{%
 \ifx #1\expandafter \@firstoftwo
 \else \expandafter \@secondoftwo
 \fi
}%
\providecommand \natexlab [1]{#1}%
\providecommand \enquote  [1]{``#1''}%
\providecommand \bibnamefont  [1]{#1}%
\providecommand \bibfnamefont [1]{#1}%
\providecommand \citenamefont [1]{#1}%
\providecommand \href@noop [0]{\@secondoftwo}%
\providecommand \href [0]{\begingroup \@sanitize@url \@href}%
\providecommand \@href[1]{\@@startlink{#1}\@@href}%
\providecommand \@@href[1]{\endgroup#1\@@endlink}%
\providecommand \@sanitize@url [0]{\catcode `\\12\catcode `\$12\catcode
  `\&12\catcode `\#12\catcode `\^12\catcode `\_12\catcode `\%12\relax}%
\providecommand \@@startlink[1]{}%
\providecommand \@@endlink[0]{}%
\providecommand \url  [0]{\begingroup\@sanitize@url \@url }%
\providecommand \@url [1]{\endgroup\@href {#1}{\urlprefix }}%
\providecommand \urlprefix  [0]{URL }%
\providecommand \Eprint [0]{\href }%
\providecommand \doibase [0]{http://dx.doi.org/}%
\providecommand \selectlanguage [0]{\@gobble}%
\providecommand \bibinfo  [0]{\@secondoftwo}%
\providecommand \bibfield  [0]{\@secondoftwo}%
\providecommand \translation [1]{[#1]}%
\providecommand \BibitemOpen [0]{}%
\providecommand \bibitemStop [0]{}%
\providecommand \bibitemNoStop [0]{.\EOS\space}%
\providecommand \EOS [0]{\spacefactor3000\relax}%
\providecommand \BibitemShut  [1]{\csname bibitem#1\endcsname}%
\let\auto@bib@innerbib\@empty
\bibitem [{\citenamefont {Deco}\ \emph {et~al.}(2009)\citenamefont {Deco},
  \citenamefont {Jirsa}, \citenamefont {McIntosh}, \citenamefont {Sporns},\
  and\ \citenamefont {Kötter}}]{deco2009}%
  \BibitemOpen
  \bibfield  {author} {\bibinfo {author} {\bibfnamefont {G.}~\bibnamefont
  {Deco}}, \bibinfo {author} {\bibfnamefont {V.}~\bibnamefont {Jirsa}},
  \bibinfo {author} {\bibfnamefont {A.~R.}\ \bibnamefont {McIntosh}}, \bibinfo
  {author} {\bibfnamefont {O.}~\bibnamefont {Sporns}}, \ and\ \bibinfo {author}
  {\bibfnamefont {R.}~\bibnamefont {Kötter}},\ }\href {\doibase
  10.1073/pnas.0901831106} {\bibfield  {journal} {\bibinfo  {journal}
  {Proceedings of the National Academy of Sciences}\ }\textbf {\bibinfo
  {volume} {106}},\ \bibinfo {pages} {10302} (\bibinfo {year}
  {2009})}\BibitemShut {NoStop}%
\bibitem [{\citenamefont {Dhamala}\ \emph {et~al.}(2004)\citenamefont
  {Dhamala}, \citenamefont {Jirsa},\ and\ \citenamefont {Ding}}]{dhamala2004}%
  \BibitemOpen
  \bibfield  {author} {\bibinfo {author} {\bibfnamefont {M.}~\bibnamefont
  {Dhamala}}, \bibinfo {author} {\bibfnamefont {V.~K.}\ \bibnamefont {Jirsa}},
  \ and\ \bibinfo {author} {\bibfnamefont {M.}~\bibnamefont {Ding}},\ }\href
  {\doibase 10.1103/PhysRevLett.92.074104} {\bibfield  {journal} {\bibinfo
  {journal} {Physical Review Letters}\ }\textbf {\bibinfo {volume} {92}},\
  \bibinfo {pages} {074104} (\bibinfo {year} {2004})}\BibitemShut {NoStop}%
\bibitem [{\citenamefont {Erneux}\ and\ \citenamefont
  {Grasman}(2008)}]{erneux2008}%
  \BibitemOpen
  \bibfield  {author} {\bibinfo {author} {\bibfnamefont {T.}~\bibnamefont
  {Erneux}}\ and\ \bibinfo {author} {\bibfnamefont {J.}~\bibnamefont
  {Grasman}},\ }\href {\doibase 10.1103/PhysRevE.78.026209} {\bibfield
  {journal} {\bibinfo  {journal} {Physical Review E}\ }\textbf {\bibinfo
  {volume} {78}},\ \bibinfo {pages} {026209} (\bibinfo {year}
  {2008})}\BibitemShut {NoStop}%
\bibitem [{\citenamefont {Takamatsu}\ \emph {et~al.}(2000)\citenamefont
  {Takamatsu}, \citenamefont {Fujii},\ and\ \citenamefont
  {Endo}}]{takamatsu2000}%
  \BibitemOpen
  \bibfield  {author} {\bibinfo {author} {\bibfnamefont {A.}~\bibnamefont
  {Takamatsu}}, \bibinfo {author} {\bibfnamefont {T.}~\bibnamefont {Fujii}}, \
  and\ \bibinfo {author} {\bibfnamefont {I.}~\bibnamefont {Endo}},\ }\href
  {\doibase 10.1103/PhysRevLett.85.2026} {\bibfield  {journal} {\bibinfo
  {journal} {Physical Review Letters}\ }\textbf {\bibinfo {volume} {85}},\
  \bibinfo {pages} {2026} (\bibinfo {year} {2000})}\BibitemShut {NoStop}%
\bibitem [{\citenamefont {Takamatsu}(2006)}]{takamatsu2006}%
  \BibitemOpen
  \bibfield  {author} {\bibinfo {author} {\bibfnamefont {A.}~\bibnamefont
  {Takamatsu}},\ }\href {\doibase 10.1016/j.physd.2006.09.001} {\bibfield
  {journal} {\bibinfo  {journal} {Physica D: Nonlinear Phenomena}\ }\textbf
  {\bibinfo {volume} {223}},\ \bibinfo {pages} {180} (\bibinfo {year}
  {2006})}\BibitemShut {NoStop}%
\bibitem [{\citenamefont {Rossoni}\ \emph {et~al.}(2005)\citenamefont
  {Rossoni}, \citenamefont {Chen}, \citenamefont {Ding},\ and\ \citenamefont
  {Feng}}]{rossoni2005}%
  \BibitemOpen
  \bibfield  {author} {\bibinfo {author} {\bibfnamefont {E.}~\bibnamefont
  {Rossoni}}, \bibinfo {author} {\bibfnamefont {Y.}~\bibnamefont {Chen}},
  \bibinfo {author} {\bibfnamefont {M.}~\bibnamefont {Ding}}, \ and\ \bibinfo
  {author} {\bibfnamefont {J.}~\bibnamefont {Feng}},\ }\href {\doibase
  10.1103/PhysRevE.71.061904} {\bibfield  {journal} {\bibinfo  {journal}
  {Physical Review E}\ }\textbf {\bibinfo {volume} {71}},\ \bibinfo {pages}
  {061904} (\bibinfo {year} {2005})}\BibitemShut {NoStop}%
\bibitem [{\citenamefont {Kauffman}\ and\ \citenamefont
  {Wille}(1975)}]{kauffman1975}%
  \BibitemOpen
  \bibfield  {author} {\bibinfo {author} {\bibfnamefont {S.}~\bibnamefont
  {Kauffman}}\ and\ \bibinfo {author} {\bibfnamefont {J.~J.}\ \bibnamefont
  {Wille}},\ }\href {\doibase 10.1016/S0022-5193(75)80108-1} {\bibfield
  {journal} {\bibinfo  {journal} {Journal of Theoretical Biology}\ }\textbf
  {\bibinfo {volume} {55}},\ \bibinfo {pages} {47} (\bibinfo {year}
  {1975})}\BibitemShut {NoStop}%
\bibitem [{\citenamefont {Bregni}(2002)}]{bregni2002}%
  \BibitemOpen
  \bibfield  {author} {\bibinfo {author} {\bibfnamefont {S.}~\bibnamefont
  {Bregni}},\ }\href@noop {} {\emph {\bibinfo {title} {Synchronization of
  {Digital} {Telecommunications} {Networks}}}}\ (\bibinfo  {publisher}
  {Wiley},\ \bibinfo {year} {2002})\BibitemShut {NoStop}%
\bibitem [{\citenamefont {Stephan}(1986)}]{stephan1986}%
  \BibitemOpen
  \bibfield  {author} {\bibinfo {author} {\bibfnamefont {K.}~\bibnamefont
  {Stephan}},\ }\href {\doibase 10.1109/TMTT.1986.1133491} {\bibfield
  {journal} {\bibinfo  {journal} {IEEE Transactions on Microwave Theory and
  Techniques}\ }\textbf {\bibinfo {volume} {34}},\ \bibinfo {pages} {1017}
  (\bibinfo {year} {1986})}\BibitemShut {NoStop}%
\bibitem [{\citenamefont {Hoppensteadt}\ and\ \citenamefont
  {Izhikevich}(2001)}]{hoppensteadt2001}%
  \BibitemOpen
  \bibfield  {author} {\bibinfo {author} {\bibfnamefont {F.}~\bibnamefont
  {Hoppensteadt}}\ and\ \bibinfo {author} {\bibfnamefont {E.}~\bibnamefont
  {Izhikevich}},\ }\href {\doibase 10.1109/81.904877} {\bibfield  {journal}
  {\bibinfo  {journal} {IEEE Transactions on Circuits and Systems I:
  Fundamental Theory and Applications}\ }\textbf {\bibinfo {volume} {48}},\
  \bibinfo {pages} {133} (\bibinfo {year} {2001})}\BibitemShut {NoStop}%
\bibitem [{\citenamefont {Wünsche}\ \emph {et~al.}(2005)\citenamefont
  {Wünsche}, \citenamefont {Bauer}, \citenamefont {Kreissl}, \citenamefont
  {Ushakov}, \citenamefont {Korneyev}, \citenamefont {Henneberger},
  \citenamefont {Wille}, \citenamefont {Erzgräber}, \citenamefont {Peil},
  \citenamefont {Elsäßer},\ and\ \citenamefont {Fischer}}]{fischer2005}%
  \BibitemOpen
  \bibfield  {author} {\bibinfo {author} {\bibfnamefont {H.-J.}\ \bibnamefont
  {Wünsche}}, \bibinfo {author} {\bibfnamefont {S.}~\bibnamefont {Bauer}},
  \bibinfo {author} {\bibfnamefont {J.}~\bibnamefont {Kreissl}}, \bibinfo
  {author} {\bibfnamefont {O.}~\bibnamefont {Ushakov}}, \bibinfo {author}
  {\bibfnamefont {N.}~\bibnamefont {Korneyev}}, \bibinfo {author}
  {\bibfnamefont {F.}~\bibnamefont {Henneberger}}, \bibinfo {author}
  {\bibfnamefont {E.}~\bibnamefont {Wille}}, \bibinfo {author} {\bibfnamefont
  {H.}~\bibnamefont {Erzgräber}}, \bibinfo {author} {\bibfnamefont
  {M.}~\bibnamefont {Peil}}, \bibinfo {author} {\bibfnamefont {W.}~\bibnamefont
  {Elsäßer}}, \ and\ \bibinfo {author} {\bibfnamefont {I.}~\bibnamefont
  {Fischer}},\ }\href {\doibase 10.1103/PhysRevLett.94.163901} {\bibfield
  {journal} {\bibinfo  {journal} {Physical Review Letters}\ }\textbf {\bibinfo
  {volume} {94}},\ \bibinfo {pages} {163901} (\bibinfo {year}
  {2005})}\BibitemShut {NoStop}%
\bibitem [{\citenamefont {Shim}\ \emph {et~al.}(2007)\citenamefont {Shim},
  \citenamefont {Imboden},\ and\ \citenamefont {Mohanty}}]{shim2007}%
  \BibitemOpen
  \bibfield  {author} {\bibinfo {author} {\bibfnamefont {S.-B.}\ \bibnamefont
  {Shim}}, \bibinfo {author} {\bibfnamefont {M.}~\bibnamefont {Imboden}}, \
  and\ \bibinfo {author} {\bibfnamefont {P.}~\bibnamefont {Mohanty}},\ }\href
  {\doibase 10.1126/science.1137307} {\bibfield  {journal} {\bibinfo  {journal}
  {Science}\ }\textbf {\bibinfo {volume} {316}},\ \bibinfo {pages} {95}
  (\bibinfo {year} {2007})}\BibitemShut {NoStop}%
\bibitem [{\citenamefont {Zhang}\ \emph {et~al.}(2012)\citenamefont {Zhang},
  \citenamefont {Wiederhecker}, \citenamefont {Manipatruni}, \citenamefont
  {Barnard}, \citenamefont {McEuen},\ and\ \citenamefont {Lipson}}]{zhang2012}%
  \BibitemOpen
  \bibfield  {author} {\bibinfo {author} {\bibfnamefont {M.}~\bibnamefont
  {Zhang}}, \bibinfo {author} {\bibfnamefont {G.~S.}\ \bibnamefont
  {Wiederhecker}}, \bibinfo {author} {\bibfnamefont {S.}~\bibnamefont
  {Manipatruni}}, \bibinfo {author} {\bibfnamefont {A.}~\bibnamefont
  {Barnard}}, \bibinfo {author} {\bibfnamefont {P.}~\bibnamefont {McEuen}}, \
  and\ \bibinfo {author} {\bibfnamefont {M.}~\bibnamefont {Lipson}},\ }\href
  {http://prl.aps.org/abstract/PRL/v109/i23/e233906} {\bibfield  {journal}
  {\bibinfo  {journal} {Physical review letters}\ }\textbf {\bibinfo {volume}
  {109}},\ \bibinfo {pages} {233906} (\bibinfo {year} {2012})}\BibitemShut
  {NoStop}%
\bibitem [{\citenamefont {Holmes}\ \emph {et~al.}(2012)\citenamefont {Holmes},
  \citenamefont {Meaney},\ and\ \citenamefont {Milburn}}]{milburn2012}%
  \BibitemOpen
  \bibfield  {author} {\bibinfo {author} {\bibfnamefont {C.~A.}\ \bibnamefont
  {Holmes}}, \bibinfo {author} {\bibfnamefont {C.~P.}\ \bibnamefont {Meaney}},
  \ and\ \bibinfo {author} {\bibfnamefont {G.~J.}\ \bibnamefont {Milburn}},\
  }\href {\doibase 10.1103/PhysRevE.85.066203} {\bibfield  {journal} {\bibinfo
  {journal} {Physical Review E}\ }\textbf {\bibinfo {volume} {85}},\ \bibinfo
  {pages} {066203} (\bibinfo {year} {2012})}\BibitemShut {NoStop}%
\bibitem [{\citenamefont {León~Aveleyra}\ \emph {et~al.}(2014)\citenamefont
  {León~Aveleyra}, \citenamefont {Holmes},\ and\ \citenamefont
  {Milburn}}]{milburn2014}%
  \BibitemOpen
  \bibfield  {author} {\bibinfo {author} {\bibfnamefont {G.}~\bibnamefont
  {León~Aveleyra}}, \bibinfo {author} {\bibfnamefont {C.~A.}\ \bibnamefont
  {Holmes}}, \ and\ \bibinfo {author} {\bibfnamefont {G.~J.}\ \bibnamefont
  {Milburn}},\ }\href {\doibase 10.1103/PhysRevE.89.062912} {\bibfield
  {journal} {\bibinfo  {journal} {Physical Review E}\ }\textbf {\bibinfo
  {volume} {89}},\ \bibinfo {pages} {062912} (\bibinfo {year}
  {2014})}\BibitemShut {NoStop}%
\bibitem [{\citenamefont {Cross}\ \emph {et~al.}(2004)\citenamefont {Cross},
  \citenamefont {Zumdieck}, \citenamefont {Lifshitz},\ and\ \citenamefont
  {Rogers}}]{cross2004}%
  \BibitemOpen
  \bibfield  {author} {\bibinfo {author} {\bibfnamefont {M.~C.}\ \bibnamefont
  {Cross}}, \bibinfo {author} {\bibfnamefont {A.}~\bibnamefont {Zumdieck}},
  \bibinfo {author} {\bibfnamefont {R.}~\bibnamefont {Lifshitz}}, \ and\
  \bibinfo {author} {\bibfnamefont {J.~L.}\ \bibnamefont {Rogers}},\ }\href
  {\doibase 10.1103/PhysRevLett.93.224101} {\bibfield  {journal} {\bibinfo
  {journal} {Physical Review Letters}\ }\textbf {\bibinfo {volume} {93}},\
  \bibinfo {pages} {224101} (\bibinfo {year} {2004})}\BibitemShut {NoStop}%
\bibitem [{\citenamefont {Zhang}\ \emph {et~al.}(2015)\citenamefont {Zhang},
  \citenamefont {Shah}, \citenamefont {Cardenas},\ and\ \citenamefont
  {Lipson}}]{zhang2015}%
  \BibitemOpen
  \bibfield  {author} {\bibinfo {author} {\bibfnamefont {M.}~\bibnamefont
  {Zhang}}, \bibinfo {author} {\bibfnamefont {S.}~\bibnamefont {Shah}},
  \bibinfo {author} {\bibfnamefont {J.}~\bibnamefont {Cardenas}}, \ and\
  \bibinfo {author} {\bibfnamefont {M.}~\bibnamefont {Lipson}},\ }\href
  {\doibase 10.1103/PhysRevLett.115.163902} {\bibfield  {journal} {\bibinfo
  {journal} {Physical Review Letters}\ }\textbf {\bibinfo {volume} {115}},\
  \bibinfo {pages} {163902} (\bibinfo {year} {2015})}\BibitemShut {NoStop}%
\bibitem [{\citenamefont {Bagheri}\ \emph {et~al.}(2013)\citenamefont
  {Bagheri}, \citenamefont {Poot}, \citenamefont {Fan}, \citenamefont
  {Marquardt},\ and\ \citenamefont {Tang}}]{bagheri2013}%
  \BibitemOpen
  \bibfield  {author} {\bibinfo {author} {\bibfnamefont {M.}~\bibnamefont
  {Bagheri}}, \bibinfo {author} {\bibfnamefont {M.}~\bibnamefont {Poot}},
  \bibinfo {author} {\bibfnamefont {L.}~\bibnamefont {Fan}}, \bibinfo {author}
  {\bibfnamefont {F.}~\bibnamefont {Marquardt}}, \ and\ \bibinfo {author}
  {\bibfnamefont {H.~X.}\ \bibnamefont {Tang}},\ }\href {\doibase
  10.1103/PhysRevLett.111.213902} {\bibfield  {journal} {\bibinfo  {journal}
  {Physical Review Letters}\ }\textbf {\bibinfo {volume} {111}},\ \bibinfo
  {pages} {213902} (\bibinfo {year} {2013})}\BibitemShut {NoStop}%
\bibitem [{\citenamefont {Matheny}\ \emph {et~al.}(2014)\citenamefont
  {Matheny}, \citenamefont {Grau}, \citenamefont {Villanueva}, \citenamefont
  {Karabalin}, \citenamefont {Cross},\ and\ \citenamefont
  {Roukes}}]{matheny2014}%
  \BibitemOpen
  \bibfield  {author} {\bibinfo {author} {\bibfnamefont {M.~H.}\ \bibnamefont
  {Matheny}}, \bibinfo {author} {\bibfnamefont {M.}~\bibnamefont {Grau}},
  \bibinfo {author} {\bibfnamefont {L.~G.}\ \bibnamefont {Villanueva}},
  \bibinfo {author} {\bibfnamefont {R.~B.}\ \bibnamefont {Karabalin}}, \bibinfo
  {author} {\bibfnamefont {M.}~\bibnamefont {Cross}}, \ and\ \bibinfo {author}
  {\bibfnamefont {M.~L.}\ \bibnamefont {Roukes}},\ }\href {\doibase
  10.1103/PhysRevLett.112.014101} {\bibfield  {journal} {\bibinfo  {journal}
  {Physical Review Letters}\ }\textbf {\bibinfo {volume} {112}},\ \bibinfo
  {pages} {014101} (\bibinfo {year} {2014})}\BibitemShut {NoStop}%
\bibitem [{\citenamefont {Agrawal}\ \emph {et~al.}(2013)\citenamefont
  {Agrawal}, \citenamefont {Woodhouse},\ and\ \citenamefont
  {Seshia}}]{seshia2013}%
  \BibitemOpen
  \bibfield  {author} {\bibinfo {author} {\bibfnamefont {D.~K.}\ \bibnamefont
  {Agrawal}}, \bibinfo {author} {\bibfnamefont {J.}~\bibnamefont {Woodhouse}},
  \ and\ \bibinfo {author} {\bibfnamefont {A.~A.}\ \bibnamefont {Seshia}},\
  }\href {\doibase 10.1103/PhysRevLett.111.084101} {\bibfield  {journal}
  {\bibinfo  {journal} {Physical Review Letters}\ }\textbf {\bibinfo {volume}
  {111}},\ \bibinfo {pages} {084101} (\bibinfo {year} {2013})}\BibitemShut
  {NoStop}%
\bibitem [{\citenamefont {Heinrich}\ \emph {et~al.}(2011)\citenamefont
  {Heinrich}, \citenamefont {Ludwig}, \citenamefont {Qian}, \citenamefont
  {Kubala},\ and\ \citenamefont {Marquardt}}]{heinrich2011}%
  \BibitemOpen
  \bibfield  {author} {\bibinfo {author} {\bibfnamefont {G.}~\bibnamefont
  {Heinrich}}, \bibinfo {author} {\bibfnamefont {M.}~\bibnamefont {Ludwig}},
  \bibinfo {author} {\bibfnamefont {J.}~\bibnamefont {Qian}}, \bibinfo {author}
  {\bibfnamefont {B.}~\bibnamefont {Kubala}}, \ and\ \bibinfo {author}
  {\bibfnamefont {F.}~\bibnamefont {Marquardt}},\ }\href {\doibase
  10.1103/PhysRevLett.107.043603} {\bibfield  {journal} {\bibinfo  {journal}
  {Physical Review Letters}\ }\textbf {\bibinfo {volume} {107}},\ \bibinfo
  {pages} {043603} (\bibinfo {year} {2011})}\BibitemShut {NoStop}%
\bibitem [{\citenamefont {Choi}\ \emph {et~al.}(2000)\citenamefont {Choi},
  \citenamefont {Kim}, \citenamefont {Kim},\ and\ \citenamefont
  {Hong}}]{choi2000}%
  \BibitemOpen
  \bibfield  {author} {\bibinfo {author} {\bibfnamefont {M.~Y.}\ \bibnamefont
  {Choi}}, \bibinfo {author} {\bibfnamefont {H.~J.}\ \bibnamefont {Kim}},
  \bibinfo {author} {\bibfnamefont {D.}~\bibnamefont {Kim}}, \ and\ \bibinfo
  {author} {\bibfnamefont {H.}~\bibnamefont {Hong}},\ }\href {\doibase
  10.1103/PhysRevE.61.371} {\bibfield  {journal} {\bibinfo  {journal} {Physical
  Review E}\ }\textbf {\bibinfo {volume} {61}},\ \bibinfo {pages} {371}
  (\bibinfo {year} {2000})}\BibitemShut {NoStop}%
\bibitem [{\citenamefont {Kim}\ \emph {et~al.}(1997)\citenamefont {Kim},
  \citenamefont {Park},\ and\ \citenamefont {Ryu}}]{kim1997}%
  \BibitemOpen
  \bibfield  {author} {\bibinfo {author} {\bibfnamefont {S.}~\bibnamefont
  {Kim}}, \bibinfo {author} {\bibfnamefont {S.~H.}\ \bibnamefont {Park}}, \
  and\ \bibinfo {author} {\bibfnamefont {C.~S.}\ \bibnamefont {Ryu}},\ }\href
  {\doibase 10.1103/PhysRevLett.79.2911} {\bibfield  {journal} {\bibinfo
  {journal} {Physical Review Letters}\ }\textbf {\bibinfo {volume} {79}},\
  \bibinfo {pages} {2911} (\bibinfo {year} {1997})}\BibitemShut {NoStop}%
\bibitem [{\citenamefont {Schuster}\ and\ \citenamefont
  {Wagner}(1989)}]{schuster1989}%
  \BibitemOpen
  \bibfield  {author} {\bibinfo {author} {\bibfnamefont {H.~G.}\ \bibnamefont
  {Schuster}}\ and\ \bibinfo {author} {\bibfnamefont {P.}~\bibnamefont
  {Wagner}},\ }\href {\doibase 10.1143/PTP.81.939} {\bibfield  {journal}
  {\bibinfo  {journal} {Progress of Theoretical Physics}\ }\textbf {\bibinfo
  {volume} {81}},\ \bibinfo {pages} {939} (\bibinfo {year} {1989})}\BibitemShut
  {NoStop}%
\bibitem [{\citenamefont {Shah}\ \emph {et~al.}(2015)\citenamefont {Shah},
  \citenamefont {Zhang}, \citenamefont {Rand},\ and\ \citenamefont
  {Lipson}}]{shah2015}%
  \BibitemOpen
  \bibfield  {author} {\bibinfo {author} {\bibfnamefont {S.~Y.}\ \bibnamefont
  {Shah}}, \bibinfo {author} {\bibfnamefont {M.}~\bibnamefont {Zhang}},
  \bibinfo {author} {\bibfnamefont {R.}~\bibnamefont {Rand}}, \ and\ \bibinfo
  {author} {\bibfnamefont {M.}~\bibnamefont {Lipson}},\ }\href {\doibase
  10.1103/PhysRevLett.114.113602} {\bibfield  {journal} {\bibinfo  {journal}
  {Physical Review Letters}\ }\textbf {\bibinfo {volume} {114}},\ \bibinfo
  {pages} {113602} (\bibinfo {year} {2015})}\BibitemShut {NoStop}%
\bibitem [{sup()}]{supplementary2017}%
  \BibitemOpen
  \href@noop {} {}\bibinfo {note} {See Supplemental Material at [URL], which
  includes Ref. [38], for a more detailed description of experimental setup,
  measurements, and analytical description of delayed coupling.}\BibitemShut
  {Stop}%
\bibitem [{\citenamefont {Razavi}(2004)}]{razavi2004}%
  \BibitemOpen
  \bibfield  {author} {\bibinfo {author} {\bibfnamefont {B.}~\bibnamefont
  {Razavi}},\ }\href {\doibase 10.1109/JSSC.2004.831608} {\bibfield  {journal}
  {\bibinfo  {journal} {IEEE Journal of Solid-State Circuits}\ }\textbf
  {\bibinfo {volume} {39}},\ \bibinfo {pages} {1415} (\bibinfo {year}
  {2004})}\BibitemShut {NoStop}%
\bibitem [{\citenamefont {Ikeda}\ and\ \citenamefont
  {Matsumoto}(1987)}]{matsumoto1987}%
  \BibitemOpen
  \bibfield  {author} {\bibinfo {author} {\bibfnamefont {K.}~\bibnamefont
  {Ikeda}}\ and\ \bibinfo {author} {\bibfnamefont {K.}~\bibnamefont
  {Matsumoto}},\ }\href {\doibase 10.1016/0167-2789(87)90058-3} {\bibfield
  {journal} {\bibinfo  {journal} {Physica D: Nonlinear Phenomena}\ }\textbf
  {\bibinfo {volume} {29}} (\bibinfo {year} {1987}),\
  10.1016/0167-2789(87)90058-3}\BibitemShut {NoStop}%
\bibitem [{\citenamefont {Yeung}\ and\ \citenamefont
  {Strogatz}(1999)}]{strogatz1999}%
  \BibitemOpen
  \bibfield  {author} {\bibinfo {author} {\bibfnamefont {M.~K.~S.}\
  \bibnamefont {Yeung}}\ and\ \bibinfo {author} {\bibfnamefont {S.~H.}\
  \bibnamefont {Strogatz}},\ }\href {\doibase 10.1103/PhysRevLett.82.648}
  {\bibfield  {journal} {\bibinfo  {journal} {Physical Review Letters}\
  }\textbf {\bibinfo {volume} {82}} (\bibinfo {year} {1999}),\
  10.1103/PhysRevLett.82.648}\BibitemShut {NoStop}%
\bibitem [{\citenamefont {Wirkus}\ and\ \citenamefont {Rand}(2002)}]{rand2002}%
  \BibitemOpen
  \bibfield  {author} {\bibinfo {author} {\bibfnamefont {S.}~\bibnamefont
  {Wirkus}}\ and\ \bibinfo {author} {\bibfnamefont {R.}~\bibnamefont {Rand}},\
  }\href {\doibase 10.1023/A:1020536525009} {\bibfield  {journal} {\bibinfo
  {journal} {Nonlinear Dynamics}\ }\textbf {\bibinfo {volume} {30}},\ \bibinfo
  {pages} {205} (\bibinfo {year} {2002})}\BibitemShut {NoStop}%
\bibitem [{\citenamefont {Yanchuk}(2005)}]{yanchuk2005}%
  \BibitemOpen
  \bibfield  {author} {\bibinfo {author} {\bibfnamefont {S.}~\bibnamefont
  {Yanchuk}},\ }\href {\doibase 10.1103/PhysRevE.72.036205} {\bibfield
  {journal} {\bibinfo  {journal} {Physical Review E}\ }\textbf {\bibinfo
  {volume} {72}},\ \bibinfo {pages} {036205} (\bibinfo {year}
  {2005})}\BibitemShut {NoStop}%
\bibitem [{\citenamefont {Jörg}\ \emph {et~al.}(2014)\citenamefont {Jörg},
  \citenamefont {Morelli}, \citenamefont {Ares},\ and\ \citenamefont
  {Jülicher}}]{julicher2014}%
  \BibitemOpen
  \bibfield  {author} {\bibinfo {author} {\bibfnamefont {D.~J.}\ \bibnamefont
  {Jörg}}, \bibinfo {author} {\bibfnamefont {L.~G.}\ \bibnamefont {Morelli}},
  \bibinfo {author} {\bibfnamefont {S.}~\bibnamefont {Ares}}, \ and\ \bibinfo
  {author} {\bibfnamefont {F.}~\bibnamefont {Jülicher}},\ }\href {\doibase
  10.1103/PhysRevLett.112.174101} {\bibfield  {journal} {\bibinfo  {journal}
  {Physical Review Letters}\ }\textbf {\bibinfo {volume} {112}} (\bibinfo
  {year} {2014}),\ 10.1103/PhysRevLett.112.174101}\BibitemShut {NoStop}%
\bibitem [{\citenamefont {Klinshov}\ \emph {et~al.}(2015)\citenamefont
  {Klinshov}, \citenamefont {Lücken}, \citenamefont {Shchapin}, \citenamefont
  {Nekorkin},\ and\ \citenamefont {Yanchuk}}]{yanchuk2015}%
  \BibitemOpen
  \bibfield  {author} {\bibinfo {author} {\bibfnamefont {V.}~\bibnamefont
  {Klinshov}}, \bibinfo {author} {\bibfnamefont {L.}~\bibnamefont {Lücken}},
  \bibinfo {author} {\bibfnamefont {D.}~\bibnamefont {Shchapin}}, \bibinfo
  {author} {\bibfnamefont {V.}~\bibnamefont {Nekorkin}}, \ and\ \bibinfo
  {author} {\bibfnamefont {S.}~\bibnamefont {Yanchuk}},\ }\href {\doibase
  10.1103/PhysRevLett.114.178103} {\bibfield  {journal} {\bibinfo  {journal}
  {Physical Review Letters}\ }\textbf {\bibinfo {volume} {114}} (\bibinfo
  {year} {2015}),\ 10.1103/PhysRevLett.114.178103}\BibitemShut {NoStop}%
\bibitem [{\citenamefont {Peil}\ \emph {et~al.}(2009)\citenamefont {Peil},
  \citenamefont {Jacquot}, \citenamefont {Chembo}, \citenamefont {Larger},\
  and\ \citenamefont {Erneux}}]{erneux2009}%
  \BibitemOpen
  \bibfield  {author} {\bibinfo {author} {\bibfnamefont {M.}~\bibnamefont
  {Peil}}, \bibinfo {author} {\bibfnamefont {M.}~\bibnamefont {Jacquot}},
  \bibinfo {author} {\bibfnamefont {Y.~K.}\ \bibnamefont {Chembo}}, \bibinfo
  {author} {\bibfnamefont {L.}~\bibnamefont {Larger}}, \ and\ \bibinfo {author}
  {\bibfnamefont {T.}~\bibnamefont {Erneux}},\ }\href {\doibase
  10.1103/PhysRevE.79.026208} {\bibfield  {journal} {\bibinfo  {journal}
  {Physical Review E}\ }\textbf {\bibinfo {volume} {79}} (\bibinfo {year}
  {2009}),\ 10.1103/PhysRevE.79.026208},\ \bibinfo {note} {bibtex:
  erneux2009}\BibitemShut {NoStop}%
\bibitem [{\citenamefont {Zeitler}\ \emph {et~al.}(2009)\citenamefont
  {Zeitler}, \citenamefont {Daffertshofer},\ and\ \citenamefont
  {Gielen}}]{gielen2009}%
  \BibitemOpen
  \bibfield  {author} {\bibinfo {author} {\bibfnamefont {M.}~\bibnamefont
  {Zeitler}}, \bibinfo {author} {\bibfnamefont {A.}~\bibnamefont
  {Daffertshofer}}, \ and\ \bibinfo {author} {\bibfnamefont {C.~C. A.~M.}\
  \bibnamefont {Gielen}},\ }\href {\doibase 10.1103/PhysRevE.79.065203}
  {\bibfield  {journal} {\bibinfo  {journal} {Physical Review E}\ }\textbf
  {\bibinfo {volume} {79}} (\bibinfo {year} {2009}),\
  10.1103/PhysRevE.79.065203}\BibitemShut {NoStop}%
\bibitem [{\citenamefont {Ares}\ \emph {et~al.}(2012)\citenamefont {Ares},
  \citenamefont {Morelli}, \citenamefont {Jörg}, \citenamefont {Oates},\ and\
  \citenamefont {Jülicher}}]{julicher2012}%
  \BibitemOpen
  \bibfield  {author} {\bibinfo {author} {\bibfnamefont {S.}~\bibnamefont
  {Ares}}, \bibinfo {author} {\bibfnamefont {L.~G.}\ \bibnamefont {Morelli}},
  \bibinfo {author} {\bibfnamefont {D.~J.}\ \bibnamefont {Jörg}}, \bibinfo
  {author} {\bibfnamefont {A.~C.}\ \bibnamefont {Oates}}, \ and\ \bibinfo
  {author} {\bibfnamefont {F.}~\bibnamefont {Jülicher}},\ }\href {\doibase
  10.1103/PhysRevLett.108.204101} {\bibfield  {journal} {\bibinfo  {journal}
  {Physical Review Letters}\ }\textbf {\bibinfo {volume} {108}} (\bibinfo
  {year} {2012}),\ 10.1103/PhysRevLett.108.204101}\BibitemShut {NoStop}%
\bibitem [{\citenamefont {Lee}\ \emph {et~al.}(2013)\citenamefont {Lee},
  \citenamefont {Hirose}, \citenamefont {Hou},\ and\ \citenamefont
  {Kil}}]{lee2013}%
  \BibitemOpen
  \bibfield  {author} {\bibinfo {author} {\bibfnamefont {M.}~\bibnamefont
  {Lee}}, \bibinfo {author} {\bibfnamefont {A.}~\bibnamefont {Hirose}},
  \bibinfo {author} {\bibfnamefont {Z.-G.}\ \bibnamefont {Hou}}, \ and\
  \bibinfo {author} {\bibfnamefont {R.~M.}\ \bibnamefont {Kil}},\ }\href@noop
  {} {\emph {\bibinfo {title} {Neural {Information} {Processing}: 20th
  {International} {Conference}, {ICONIP} 2013, {Daegu}, {Korea}, {November}
  3-7, 2013. {Proceedings}}}}\ (\bibinfo  {publisher} {Springer},\ \bibinfo
  {year} {2013})\BibitemShut {NoStop}%
\bibitem [{\citenamefont {Cosp}\ \emph {et~al.}(2004)\citenamefont {Cosp},
  \citenamefont {Madrenas}, \citenamefont {Alarcon}, \citenamefont {Vidal},\
  and\ \citenamefont {Villar}}]{cosp2004}%
  \BibitemOpen
  \bibfield  {author} {\bibinfo {author} {\bibfnamefont {J.}~\bibnamefont
  {Cosp}}, \bibinfo {author} {\bibfnamefont {J.}~\bibnamefont {Madrenas}},
  \bibinfo {author} {\bibfnamefont {E.}~\bibnamefont {Alarcon}}, \bibinfo
  {author} {\bibfnamefont {E.}~\bibnamefont {Vidal}}, \ and\ \bibinfo {author}
  {\bibfnamefont {G.}~\bibnamefont {Villar}},\ }\href {\doibase
  10.1109/TNN.2004.832808} {\bibfield  {journal} {\bibinfo  {journal} {IEEE
  Transactions on Neural Networks}\ }\textbf {\bibinfo {volume} {15}},\
  \bibinfo {pages} {1315} (\bibinfo {year} {2004})}\BibitemShut {NoStop}%
\end{thebibliography}%


\begin{thebibliography}{2}%
\makeatletter
\providecommand \@ifxundefined [1]{%
 \@ifx{#1\undefined}
}%
\providecommand \@ifnum [1]{%
 \ifnum #1\expandafter \@firstoftwo
 \else \expandafter \@secondoftwo
 \fi
}%
\providecommand \@ifx [1]{%
 \ifx #1\expandafter \@firstoftwo
 \else \expandafter \@secondoftwo
 \fi
}%
\providecommand \natexlab [1]{#1}%
\providecommand \enquote  [1]{``#1''}%
\providecommand \bibnamefont  [1]{#1}%
\providecommand \bibfnamefont [1]{#1}%
\providecommand \citenamefont [1]{#1}%
\providecommand \href@noop [0]{\@secondoftwo}%
\providecommand \href [0]{\begingroup \@sanitize@url \@href}%
\providecommand \@href[1]{\@@startlink{#1}\@@href}%
\providecommand \@@href[1]{\endgroup#1\@@endlink}%
\providecommand \@sanitize@url [0]{\catcode `\\12\catcode `\$12\catcode
  `\&12\catcode `\#12\catcode `\^12\catcode `\_12\catcode `\%12\relax}%
\providecommand \@@startlink[1]{}%
\providecommand \@@endlink[0]{}%
\providecommand \url  [0]{\begingroup\@sanitize@url \@url }%
\providecommand \@url [1]{\endgroup\@href {#1}{\urlprefix }}%
\providecommand \urlprefix  [0]{URL }%
\providecommand \Eprint [0]{\href }%
\providecommand \doibase [0]{http://dx.doi.org/}%
\providecommand \selectlanguage [0]{\@gobble}%
\providecommand \bibinfo  [0]{\@secondoftwo}%
\providecommand \bibfield  [0]{\@secondoftwo}%
\providecommand \translation [1]{[#1]}%
\providecommand \BibitemOpen [0]{}%
\providecommand \bibitemStop [0]{}%
\providecommand \bibitemNoStop [0]{.\EOS\space}%
\providecommand \EOS [0]{\spacefactor3000\relax}%
\providecommand \BibitemShut  [1]{\csname bibitem#1\endcsname}%
\let\auto@bib@innerbib\@empty
\bibitem [{\citenamefont {Shah}\ \emph {et~al.}(2015)\citenamefont {Shah},
  \citenamefont {Zhang}, \citenamefont {Rand},\ and\ \citenamefont
  {Lipson}}]{shah2015}%
  \BibitemOpen
  \bibfield  {author} {\bibinfo {author} {\bibfnamefont {S.~Y.}\ \bibnamefont
  {Shah}}, \bibinfo {author} {\bibfnamefont {M.}~\bibnamefont {Zhang}},
  \bibinfo {author} {\bibfnamefont {R.}~\bibnamefont {Rand}}, \ and\ \bibinfo
  {author} {\bibfnamefont {M.}~\bibnamefont {Lipson}},\ }\href {\doibase
  10.1103/PhysRevLett.114.113602} {\bibfield  {journal} {\bibinfo  {journal}
  {Physical Review Letters}\ }\textbf {\bibinfo {volume} {114}},\ \bibinfo
  {pages} {113602} (\bibinfo {year} {2015})}\BibitemShut {NoStop}%
\bibitem [{\citenamefont {Haus}(1984)}]{haus1984}%
  \BibitemOpen
  \bibfield  {author} {\bibinfo {author} {\bibfnamefont {H.~A.}\ \bibnamefont
  {Haus}},\ }\href@noop {} {\emph {\bibinfo {title} {Waves and fields in
  optoelectronics}}}\ (\bibinfo  {publisher} {Prentice Hall, Incorporated},\
  \bibinfo {year} {1984})\BibitemShut {NoStop}%
\end{thebibliography}%

\end{document}



\title{Long-range Synchronization of Nanomechanical Oscillators with Light}

\author{Shreyas Y. Shah}
\affiliation{School of Electrical and Computer Engineering, Cornell University, Ithaca, New York 14853, USA}
\affiliation{School of Electrical and Computer Engineering, Purdue University, West Lafayette, Indiana 47907, USA}
\author{Mian Zhang}
\affiliation{School of Electrical and Computer Engineering, Cornell University, Ithaca, New York 14853, USA}
\affiliation{School of Engineering and Applied Sciences, Harvard University, Cambridge, Massachusetts 02138, USA}
\author{Richard Rand}
\affiliation{Department of Mathematics, Cornell University, Ithaca, New York 14853, USA}
\affiliation{Sibley School of Mechanical and Aerospace Engineering, Cornell University, Ithaca, New York 14853, USA}
\author{Michal Lipson}
\affiliation{School of Electrical and Computer Engineering, Cornell University, Ithaca, New York 14853, USA}
\affiliation{Kavli Institute at Cornell for Nanoscale Science, Ithaca, New York 14853, USA}
\affiliation{Department of Electrical Engineering, Columbia University, New York, New York 100027, USA}





\maketitle

\section{Experimental Setup and Procedure for Delay-coupled synchronisation}

\begin{center}
\begin{figure*}
\includegraphics[scale=0.55]{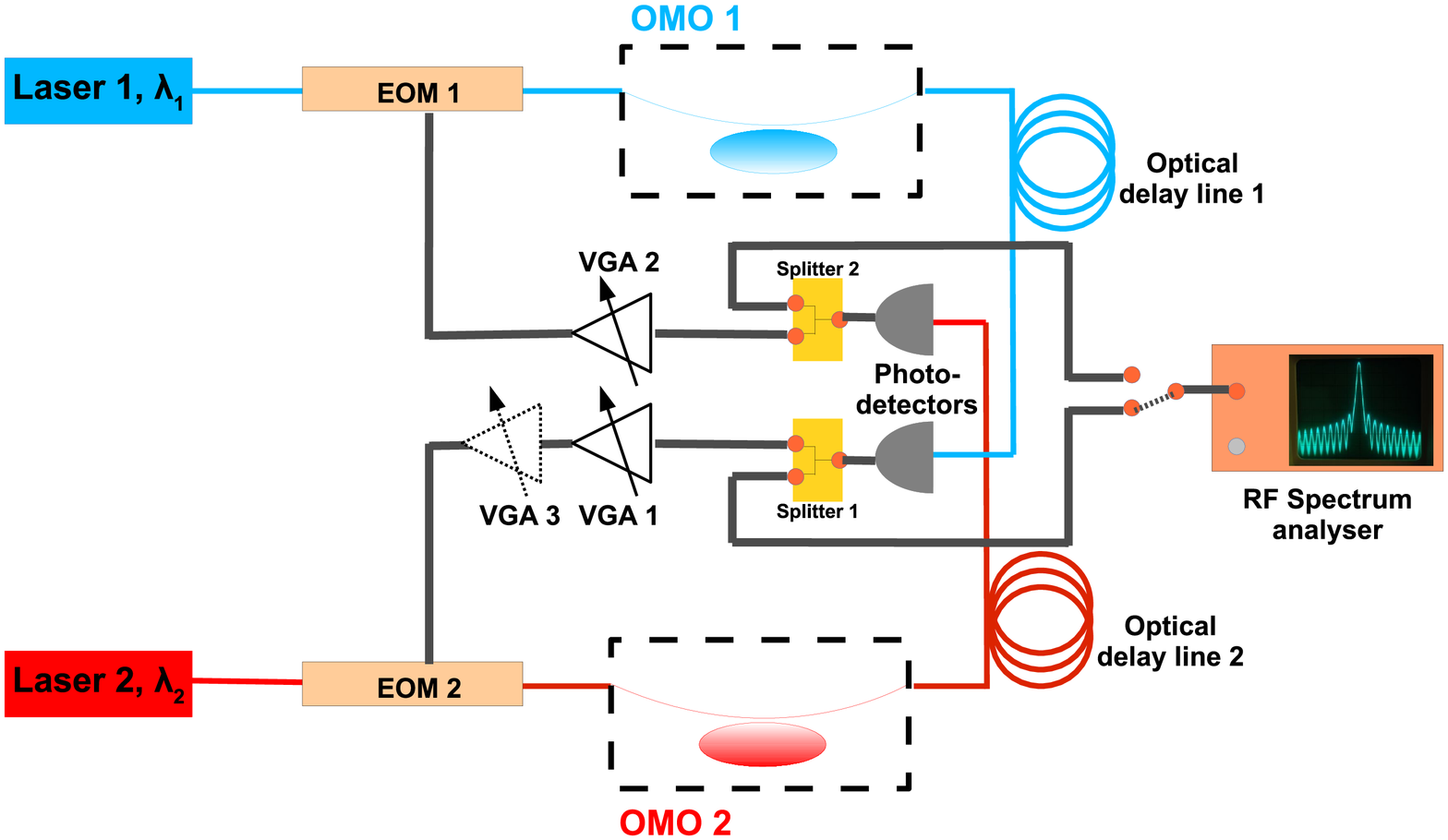}
\caption{Detailed experimental schematic}
\label{FigS1}
\end{figure*}
\end{center}

A more detailed schematic of experimental setup to synchronise two optomechanical oscillators (OMOs) is shown in Fig. \ref{FigS1}. As described in the main text, each device is driven by an independent laser tuned to be blue-side of its optical resonance. The transmitted optical signals, modulated by each OMO, travel over delay line of SMF-28 optical fibres. The RF signal generated at the photodetectors (DC filters are used to block the DC signal) at the end of optical delay lines modulate the power of the lasers driving the two OMOs via electro-optic modulators (EOM). The strengths of these modulation signals are controlled by variable-gain RF amplifiers (VGA).

The coupling strengths are primarily determined by VGA1 and VGA2. The two OMOs are first pumped into self-sustained oscillations, while keeping the gain values very low ($<-20$ dB), so that the two devices oscillate independently. VGA1 and VGA2 are controlled by the same voltage source, and have the same gain (within their specifications) as the control-voltage is varied. The synchronisation transition i.e. when the two OMOs transition from independent oscillations at different frequencies to locked oscillations at the same frequency, is seen when the gain is increased. We increase the gain in steps of $\approx$ 0.9 dB.

Half of the RF oscillation signal is tapped off at each of the photodetector for analysing with an RF spectrum analyser. Since the instrument we use only has a single input channel, we analyse and record the spectrum of each oscillator independently. Therefore, each voltage scan (as described in the previous paragraph) is performed twice, first to record the output of Splitter 1 and then to record the output of Splitter 2. The two spectra are then mathematically added using numerical software to yield a combined RF spectrum for the two OMOs. The output voltage of the voltage sources drifts over time. This is minimised by monitoring and compensating manually as needed. 

We have seen earlier \cite{shah2015} that the response of an OMO to an externally injected periodic signal is highly asymmetric with respect to the detuning between the OMO and the external signal. Therefore, an OMO is more susceptible to locking by an external signal if that signal has a higher frequency than if it has a lower freuquency. This means that, in order to observe synchronisation dynamics, it is not enough to have equal values of gain (and thereby $\kappa_{21}$ and $\kappa_{12}$).

A third amplifier VGA3, cascaded with VGA1 and controlled independently of VGA1 and VGA2, is used to differentiate between $\kappa_{21}$ and $\kappa_{12}$. The gain of VGA3 is kept fixed throughout the voltage-scan described above.

\section{Mathematical model for delayed coupling}

Each OMO can be modelled as a pair of parametrically coupled optical (Eq. \ref{EqS1}) and mechanical (Eq. \ref{EqS2}) resonators, where $a$ is the electric field strength in the optical resonator, such that $\lvert a \rvert ^2$ is the energy stored in the cavity, $\Delta_0$ is the detuning between the laser frequency and the resonance frequency of the optical cavity $\omega$, $\Gamma_{opt}$ is the optical decay rate, $\Gamma_{ext}$ is the rate at which laser power $|s|^{2}$ is coupled into the optical cavity from the tapered fiber, $G_{\text{om}}$ is the optomechanical coupling coefficient and $m_{eff}$ is the effective mass of the mechanical resonator. The rest of the parameters are described in the main text.

\begin{equation}\tag{S1}\label{EqS1}
\frac{da}{dt} = i(\Delta_{0} - G_{\text{om}}x)a - \Gamma_{opt}a + \sqrt{2\Gamma_{ex}}s
\end{equation}
\begin{equation}\tag{S2}\label{EqS2}
\frac{d^{2}x}{dt^{2}} + \Gamma_{m}\frac{dx}{dt}+\Omega_{m}^{2}x = \frac{G_{\text{om}}\lvert a \rvert ^2}{m_{eff}\omega}
\end{equation}

For the OMOs that we used in this demonstration of synchronisation, the optical decay rate $\Gamma_{opt}$ is much larger than the mechanical frequency $\Omega_m$, and Eqs. \ref{EqS1}, \ref{EqS2} can be approximated by a single equation Eq. \ref{EqS3}, where $\tau$ is the response time of the optical cavity.

\begin{equation}\tag{S3}\label{EqS3}
\frac{d^{2}x(t)}{dt^{2}} + \Gamma_{m}\frac{dx(t)}{dt}+\Omega_{m}^{2}x(t) = \frac{2G_{\text{om}}\Gamma_{ex}}{m_{eff}\omega}\frac{1}{(\Delta_{0} - G_{\text{om}}x(t-\tau))^{2}+\Gamma_{opt}^{2}}|s|^{2}
\end{equation}

The laser power $|s|^{2}$ driving one OMO is modulated, via an electro-optic modulator, by the RF oscillation signal of the other OMO, $P_{trans}$ (Eq. \ref{EqS4}) (\cite{shah2015}, Supplementary). Here, $\frac{\Gamma}{2}$ represents the strength of modulation due to $P_{trans}$.

\begin{equation}\tag{S4}\label{EqS4}
|s|^2 = |s_0|^2(1+\frac{\Gamma}{2}P_{trans})
\end{equation}

$P_{trans}$ is the RF oscillation power of the OMO, that modulates the laser power $|s|^2$. It can be shown that $P_{trans}(t) \propto x_{trans}(t)$ (Eq. \ref{eqS7}), for small enough $x_{trans}$. Substituting this in Eq. \ref{EqS4}, and combining it with \ref{EqS3}, assuming that $P_{trans}$ is delayed by T, we get Eq. \ref{EqS5}, which describes the delayed coupling between the two OMOs.

\begin{equation}\tag{S5}\label{EqS5}
\begin{split}
&\frac{d^{2}x(t)}{dt^{2}} + \Gamma_{m}\frac{dx(t)}{dt}+\Omega_{m}^{2}x(t) = F_{opt}(x(t))(1+\gamma x_{trans}(t-T)) \\
& \text{where, } F_{opt}(x(t)) = \frac{2G_{\text{om}}\Gamma_{ex}}{m_{eff}\omega}\frac{1}{(\Delta_{0} - G_{\text{om}}x(t-\tau))^{2}+\Gamma_{opt}^{2}}|s_0|^2
\end{split}
\end{equation}

\section{Relationship between $\kappa_{ij}$ and $\gamma_{ij}$}

Eq. \ref{EqS1} represents the optical field circulating within the optical cavity of the OMO. The power exiting the cavity can be described by $|s_{out}|^{2}=|s-\sqrt{2\Gamma_{ex}}a|^{2}$ \cite{haus1984}. By combining the transimpedance gain of the photodetector and the input gain of the RF spectrum analyzer into the term $D_{g}$, the power detected at the spectrum analyzer can be written as, for OMO $j$,

\begin{equation}\tag{S6}\label{eqS6}
P_{j}\left(x_j\right) = D_{g}|s|^{2}|1-\frac{2\Gamma_{ex}}{i(\Delta_{0}-g_{om}x_j)-\Gamma_{opt}}|^2
\end{equation}

Assuming $x_j$ oscillates at the frequency $\Omega_{j}$ i.e. $x = x_{0}\cos(\Omega_{j} t)$, $P_{j}$ can be approximated in terms of its spectral components, i.e. as a Fourier series Eq. \ref{eqS7}, where $D_{g}|s|^{2}(P_{0,j}, P_{1,j}, P_{2,j}, ...)$ are the power-spectral-density (PSD) values of $P_{j}$ at the frequencies $(0, \Omega_{j}, 2\Omega_{j}, ...)$. Harmonics are introduced because of the non-linear transduction between $x_j$ and $P_{j}$.

\begin{equation}\tag{S7}\label{eqS7}
P_{j} = D_{g}|s_{0,j}|^{2}\left(P_{0,j} + P_{1,j}\cos(\Omega_{j} t) + P_{2,j}\cos(2\Omega_{j} t) + ...\right)
\end{equation}

The parameter $H_{osc,j}$ from the main text can, therefore, be written as \linebreak  $H_{osc,j} = D_{g,j}|s_{0,j}|^2 P_{1,j}$. This value is directly read off the spectrum analyzer.

Similarly, when $P_j$ is modulated by the signal coming from oscillator $i$ via EOM $j$, we have, from Eqs. \ref{EqS4} and \ref{eqS7},

\begin{equation}\tag{S8}\label{eqS8}
P_{j} = D_{g}|s_{0,j}|^{2} (1+\frac{\Gamma}{2}P_{i,inj}) \left(P_{0,j} + P_{1,j}\cos(\Omega_{j} t) + P_{2,j}\cos(2\Omega_{j} t) + ...\right)
\end{equation}

Here, $P_{i,inj} \propto \left(P_{1,i}\cos(\Omega_{i} t) + P_{2,i}\cos(2\Omega_{i} t) + ...\right)$, because we block the DC portion of the signal before it arrives at the modulator. This proportionality constant can be absorbed in $\Gamma$. Therefore, $H_{inj,i} = D_{g,j}|s_{j}|^2 P_{0,j} P_{1,i} \frac{\Gamma}{2}$.
Therefore, considering $P_{1,i} \propto x_i$

\begin{equation}\tag{S9}\label{eqS9}
\kappa_{ij} = \frac{H_{in,j}}{H_{osc.j}} = \frac{\frac{\Gamma}{2} P_{1,i}}{{\nicefrac{P_{1,j}}{P_{0,j}}}} 
\end{equation}

Comparing \ref{EqS5} and \ref{eqS9}, it can be easily seen that $\kappa_{ij} \propto \gamma_{ij}$.

\bibliography{LDFS}